\documentclass[preprint,showpacs,preprintnumbers,amsmath,amssymb]{revtex4}

\newcommand{\ds}{\displaystyle}

\usepackage{graphicx}
\usepackage{dcolumn}
\usepackage{bm}

\newcommand{\be}{\begin{equation}}
\newcommand{\en}{\end{equation}}
\newcommand{\bea}{\begin{eqnarray}}
\newcommand{\ena}{\end{eqnarray}}

\begin{document}


\title{ The curvaton field and the intermediate inflationary universe model}

\author{Sergio del Campo}
 \email{sdelcamp@ucv.cl}
\affiliation{ Instituto de F\'{\i}sica, Pontificia Universidad
Cat\'{o}lica de Valpara\'{\i}so, Av. Brasil 2950, Valpara\'{\i}so,
Chile.}

\author{Ram\'on Herrera}
\email{ramon.herrera@ucv.cl} \affiliation{ Instituto de
F\'{\i}sica, Pontificia Universidad Cat\'{o}lica de
Valpara\'{\i}so, Av. Brasil 2950, Valpara\'{\i}so, Chile.}

\date{\today}

\begin{abstract}
The curvaton in an intermediate  inflationary universe model  is
studied. This study has allowed us to find some interesting
constraints on different parameters that appear in the model.
\end{abstract}

\maketitle

\section{Introduction}

An intermediate inflation model was introduced as an exact
solution for a particular scalar field potential of the type
$V(\phi)\propto \phi^{-\beta}$\cite{Barrow1}, where $\beta$ is a
free parameter. With this sort of potential, and with $\beta > 0$,
it is possible in the slow-roll approximation to have a spectrum
of density perturbations which presents a scale-invariant spectral
index $n_s=1$, i.e. the so-called Harrizon-Zel'dovich spectrum of
density perturbations, provided $\beta $ takes the value
two\cite{Barrow2}. Even though  this kind of spectrum is
disfavored by the current Wilkinson Microwave Anisotropy Probe
(WMAP) data\cite{WMAP3}, the inclusion of tensor perturbations,
which could be present at some point by inflation and parametrized
by the tensor-to-scalar ration $r$, the conclusion that $n_s \geq
1$ is allowed provided that he value of $r$ is significantly
nonzero\cite{ratio r}. In fact, in ref. \cite{Barrow3} was shown
that the combination $n_s=1$ and $r>0$ is given by a version of
the intermediate inflation model in which the scale factor varies
as $a(t)\propto e^{(t/t_o)^{2/3}}$ and the slow-roll approximation
was used.

The main motivation to study this sort of model becomes from
string/M-theory. This theory suggests that in order to have a
ghost-free action high order curvature invariant corrections to
the Einstein-Hilbert action must be proportional to the
Gauss-Bonnet (GB) term{\cite{BD}}. GB terms arise naturally as the
leading order of the $\alpha$ expansion to the low-energy string
effective action, where $\alpha$ is the inverse string
tension{\cite{KM}}. This kind of theory has been applied to
possible resolution of the initial singularity
problem{\cite{ART}}, to the study of Black-Hole solutions{\cite{
Varios1}}, accelerated cosmological solutions{\cite{ Varios2}}. In
particular , very recently, it has been found{\cite{Sanyal}} that
for a dark energy model the GB interaction in four dimensions with
a dynamical dilatonic scalar field coupling leads to a solution of
the form $a = a_0 \exp{A t^{f}}$,  where the universe starts
evolving with a decelerated exponential expansion. Here, the
constant $A$ becomes given by $A= \frac{2}{\kappa n}$ and
$f=\frac{1}{2}$, with $\kappa^2 = 8 \pi G$ and $n$ is a constant.
In this way, the idea that inflation , or specifically,
intermediate inflation, comes from an effective theory at low
dimension of a more fundamental string theory is in itself very
appealing.

The characteristic of the scalar potential $V(\phi)$ in this kind
of model it does not present a minimum, so that the usual
mechanism introduced to bring inflation to an end becomes useless.
In fact, the standard mechanism is described by the stage of
oscillations of the scalar field which is a essential part of the
so called reheating mechanism, where most of the matter and
radiation of the universe was created, via the decay of the
inflaton field, while the temperature grows in many orders of
magnitude. It is at this point where the Big-Bang universe is
recovered. Here, the reheating temperature, the temperature
associated to the temperature of the universe when the Big-Bang
model begins, is of particular interest. In this epoch the
radiation domination begins, in which there exist a number of
particles of different kinds.

The stage of oscillation of the scalar field is a essential part
for the standard mechanism of reheating. Therefore a minimum in
the inflaton potential is something crucial for the reheating
mechanism. However, there are models where such a minimum does not
exist, and thus the standard mechanism of reheating does not
work\cite{Kofman}. These models are known in the literature like
non-oscillating models, or simple NO models\cite{Felder}. One of
the mechanism of reheating in this kind of models is the
introduction of the curvaton field\cite{Mollerach,ref1u}. Here,
the decay of the curvaton field into conventional matter offers an
efficient mechanism of reheating, and its field has the property
whose energy density is not diluted during inflation so that the
curvaton may be responsible for some or all the matter content of
the universe at present. On the other hand, this field may also be
the responsible for explaining the observed large scale structure
of the universe.

In the context of intermediate inflation we would like to
introduce the curvaton field as a mechanism to bring intermediate
inflation to an end. Therefore, the main goal of the present paper
is to implement the curvaton field into the intermediate
inflationary scenario and see what consequences we may extract.

The outline of the paper goes as follow: in section II we give a
brief description of the intermediate inflationary scenario. In
section III the curvaton field is described in the kinetic epoch.
Section IV describes the curvaton decay after its domination.
Section V describe the decay of the curvaton field before it
dominates. Section VI studies the consequences of the
gravitational waves. At the end, section VII exhibits our
conclusions.

\section{ Intermediate inflation  Model \label{secti}}

In order to describe intermediate inflationary universe models we
start with the following field equations in a flat
Friedmann-Robertson-Walker background

\begin{equation}
 3\;H^2=\frac{\dot{\phi}^2}{2}+V(\phi)\label{key_02},
\end{equation}
and \begin{equation} \ddot{\phi}+3H\dot{\phi}=-\frac{\partial
V(\phi)}{\partial\phi},\label{3}
\end{equation}
where  $a$ is a scale factor, $H \equiv \dot{a}/a$ is the Hubble
factor, $\phi$ is the standard inflaton field and $V(\phi)$ is the
effective scalar potential. Dots here mean derivatives with
respect to the cosmological time, $t$, and we use  units in which
$8\pi G=8\pi/m_p^2=c=\hbar=1$ ($m_p$ is the Planck mass).

Exact solutions can  be found for intermediate inflationary
universe model where  the scale factor $a(t)$ expands as
\begin{equation}
a(t)=\exp(\,A\,t^{f}).\label{at}
\end{equation}
Here $f$ is a constant parameter with range $0<f<1$ and $A$ is a
positive constant.

From Eqs.(\ref{key_02}), (\ref{3}) and (\ref{at}) the expressions
for the scalar potential, $V(\phi)$ and the scalar field,
$\phi(t)$, become
\begin{equation}
V(\phi)=\frac{8A^2}{(\beta+4)^2}\left[\frac{\phi}{(2A\beta)^{1/2}}\right]^{-\beta}
\left[6-\frac{\beta^2}{\phi^2}\right],\label{PP}
\end{equation}
 and
\begin{equation}
\phi(t)=(2\;A\beta\;t^{f})^{1/2},\label{pat}
\end{equation}
respectively.  Here, the parameter $\beta$ is defined by $\beta
\equiv 4(f^{-1}-1)$.

The Hubble parameter as a function of the inflaton field, $\phi$,
becomes
\begin{equation}
H(\phi)=A\;f\;(2\;A\beta)^{\beta/4}\;\phi^{-\beta/2}.\label{HH}
\end{equation}

The form for the scale factor $a$ expressed by Eq.(\ref{at}) also
arises when we solve the field equations in the slow roll
approximation, where a simple power law scalar potential is
considered
\begin{equation}
V(\phi)=\frac{48\;A^2}{(\beta+4)^2}(2A\beta)^{\beta/2}\;\phi^{-\beta}.
\label{PP1}
\end{equation}
Note that this kind of potential does not present a minimum. Also,
the solutions for $\phi(t)$ and $H(\phi)$ obtained with this
potential in the slow roll approximation are identical to those
obtained in the exact solution, expressed by Eqs.(\ref{pat}) and
(\ref{HH}).

The slow roll parameters $\varepsilon$ and $\eta$ are defined by
$\varepsilon=\frac{V'^2}{2V^2}$ and $\eta=V''/V$, respectively,
where the prime denotes derivative with respect to the inflaton
field $\phi$. In our case they reduced to
$\varepsilon=\frac{\beta^2}{2\;\phi^2}$ and
$\eta=\frac{\beta(\beta+1)}{\phi^2}$, and its  ratio,
$\varepsilon/\eta$ becomes
$\varepsilon/\eta=\frac{1}{2}\left(\frac{\beta}{\beta+1}\right)$.
Note that  $\eta$ is always larger than $\varepsilon$. Since
$\beta$ is positive, $\eta$ reaches  unity before $\varepsilon$
does. In this way, we may establish that the end of  inflation is
governed by the condition $\eta=1$ more then $\varepsilon=1$, from
which we get, at the end of inflation  $\phi_e^2 =
\beta\,(\beta+1)$, for the inflaton field. From here on, the
subscript $e$  is used to denote the end of the inflationary
period.

\section{The curvaton field during the kinetic epoch \label{sectii}}

Neglecting the term $\ds \frac{\partial V(\phi)}{\partial \phi}$
when compared with the friction term $3H\dot{\phi}$ in the field
Eq. ({\ref{3}}), the model enters to a new period which is called
the `kinetic epoch' or `kination'. In the following we will use
the subscript (or superscript)`k' to label different quantities at
the beginning of this epoch. Note that during the kination epoch
we have that $\dot{\phi}^2/2>V(\phi)$ which could be seen as a
stiff fluid since the relation between the pressure $P_\phi$ and
the energy density $\rho_\phi$, corresponds to $P_\phi=\rho_\phi$.

In the kinetic epoch the field equations (\ref{key_02}) and
(\ref{3}) becomes $
 3\;H^2\,=\frac{\dot{\phi}^2}{2} $ and $
\ddot{\phi}+3H\dot{\phi}=0$ where the latter equation could be
solved and gives $\dot{\phi}=\dot{\phi}_k \left ({\frac{
a_k}{a}}\right )^3$. This expression yields to
\begin{equation}
\rho_{\phi}(a)=\rho_{\phi}^{k}\left (\frac{a_{k}}{a}\right )^6
\label{rho},
\end{equation}
and the Hubble parameter becomes
\begin{equation}
H(a)=H=H_{k}\left (\frac{a_k}{a}\right )^3
  \label{h},
\end{equation}
where
$H_{k}^2=\frac{\rho_{\phi}^{k}}{3}\simeq\frac{\dot{\phi}_k^2}{6} $
is the value of the Hubble parameter at the beginning of the
kinetic epoch.

The curvaton field obeys the Klein-Gordon equation and we will
assume that the scalar potential associated to this field is given
by $ U(\sigma)=\frac{m^2\sigma^2}{2}$, where $m$ is the curvaton
mass.

Firstly, we assume that the energy density associated to the
inflaton field, $\rho_{\phi}$, is the dominant component when
compared with the curvaton energy density, $\rho_\sigma$.
Secondly, the curvaton field oscillates around the minimum of its
effective potential $U(\sigma)$. During the kinetic epoch the
universe remains inflaton-dominated where the curvaton  density
evolves as a non-relativistic matter, i.e.
$\rho_\sigma\propto\,a^{-3}$. The final stage corresponds to the
decay of the curvaton field into radiation and thus the standard
Big-Bang cosmology is recovered.

During the inflationary regime it is assumed that the  curvaton
field is effectively massless \cite{dimo,postma,ureña,cdch}. In
the same period the curvaton rolls down its potential until its
kinetic energy is depleted by the exponential expansion. Its
kinetic energy has almost vanished, and it becomes frozen. The
curvaton field assumes roughly a constant value, $\sigma_*\approx
\sigma_e$. Here, the subscript $``*''$ refers to the epoch when
the cosmological scale exit the horizon.

The hypothesis assumed here is that during the kinetic epoch  the
Hubble parameter decreases so that its value is comparable with
the curvaton mass, i.e. $m\simeq H$ ( at this stage, the curvaton
field becomes effectively massive). From Eq.(\ref{h}), we obtain
\begin{equation}
\frac{m}{H_{k}}=\left (\frac{a_k}{a_m}\right )^3,\label{mh}
\end{equation}
where the subscript `m' stands for quantities at the time when the
curvaton mass, $m$, is of the order of $H$ during the kinetic
epoch.

In order to prevent a period of curvaton-driven inflation the
universe must still be dominated by the inflaton field, i.e.
$\rho_{\phi}|_{a_m}=\rho_{\phi}^{m}\gg\rho_{\sigma}(
\sim\,U(\sigma_e)\simeq\,U(\sigma_*))$.  This inequality allows us
to find a constraint on the values of the curvaton field
$\sigma_*$. At the moment when $H\simeq m$, we get that
\begin{equation}
\frac{m^2\sigma_*^2}{2\rho_\phi^{m}}\ll\,\,1\,,\label{pot}
\end{equation} which implies that the curvaton field $\sigma_*$ satisfies the
constraint $\sigma_*^2\ll 6$, where we have used $ \rho_\phi^m
=\rho_\phi^k \left(\frac{a_k}{a_m}\right)^6=\rho_\phi^k\,\left
(\frac{m}{H_k}\right )^2$.

At the end of inflation, the ratio between the potential energies
becomes
\begin{equation}
\frac{U_e}{V_e}=\frac{m^2\sigma_*^2}{6 H_e^2}<\,1 \label{u},
\end{equation}
and, in this way, the curvaton energy becomes subdominant at the
end of inflation. The curvaton mass  should obey the constraint
%
\begin{equation}
m^2<\,H_e^2=
\frac{16\;A^2}{(\beta+4)^2}\;\left[\frac{2\,A}{\beta+1}\right]^{\beta/2},
\label{one}
\end{equation}
imposed by the fact that the curvaton field must be effectively
massless during the inflationary era, and thus, $m< H_e$.  In the
latter expression we have used the relation $ V_e=\frac{48
A^2}{(\beta+4)^2}(2A\beta)^{\beta/2}\phi_e^{-\beta}$.

At the time when the mass of the  curvaton field becomes
important, i.e. when $m\simeq H$, its energy density decays like a
non-relativistic matter in the form $ \rho_\sigma=
\frac{m^2\sigma_*^2}{2}\frac{a_m^3}{a^3}$, since their potential
and kinetic energy densities are comparable due to the curvaton is
undergoing quasi-harmonic oscillations.

\section{Curvaton Decay After Domination\label{sectiv}}

The decay of the curvaton field may occur in two possible
different scenarios. Firstly, if the curvaton field comes to
dominate the cosmic expansion (i.e. $\rho_\sigma>\rho_\phi$),
then, there must be a moment in which the inflaton and curvaton
energy densities become equals. Let us assume that this happen
when $a=a_{eq}$, then, from Eqs.(\ref{rho}), (\ref{h}) and using
that $\rho_\sigma\propto\,a^{-3}$ we get

\begin{equation}
\left.\frac{\rho_\sigma}{\rho_\phi}\right|_{a=a_{eq}}=\frac{m^2\sigma_*^2}{2}\frac{a_m^3\;a_{eq}^3}{a_k^6\;\rho_\phi^k}
=\frac{m^2\sigma_*^2 a_m^3 a_{eq}^3}{6\;H_k^2 \; a_k^6} =
1\label{equili},
\end{equation}
which yields to
$H_k\left(\frac{a_k}{a_{eq}}\right)^3=\frac{m\sigma_*^2}{6}$,
where we have used the relation $3\,H_k^2=\rho_\phi^k$ together
with Eq.(\ref{mh}).

From Eqs.(\ref{h}), (\ref{mh}) and (\ref{equili}), we  write a
relation for the Hubble parameter, $H(a_{eq})=H_{eq}$, in terms of
the curvaton parameters
\begin{eqnarray}
H_{eq}&=&
H_{k}\left(\frac{a_k}{a_{eq}}\right)^3=\frac{m\;\sigma_*^2}{6}.\label{heq}
\end{eqnarray}

Since the decay parameter $\Gamma_\sigma$ is constrained by
nucleosynthesis, it is required that the curvaton field decays
before nucleosynthesis, which means $H_{nucl}\sim 10^{-40}<
\Gamma_\sigma$ (in units of Planck mass $m_p$). We also require
that the curvaton decay occurs after domination, i.e. $\rho_\sigma
> \rho_\phi$, and also for $\Gamma_\sigma < H_{eq}$. Thus, we get a
constraint on the decay parameter $\Gamma_\sigma$, which is given
by
\begin{equation}
10^{-40}<\Gamma_{\sigma}<\frac{m\;\sigma_*^2}{6} .\label{gamm1}
\end{equation}

It is interesting to find constraints on the parameters appearing
in our model by studying the scalar perturbations related to the
curvaton field $\sigma$. In general, we may say that the curvaton
field creates the curvature perturbations in two separate stages.
In the first stage, the quantum fluctuations during inflation are
converted to classical perturbations characterized by a flat
spectrum at horizon exit. Then, in the second period (at the time
after inflation), the perturbations are converted into curvature
perturbations. Differently, to the usual mechanism, the generation
of curvature perturbations by the curvaton field require no
assumptions about the nature of inflation, except by the
requirement that the Hubble parameter remains practically
constant.

During the time in which the fluctuations are inside the horizon,
they obey the same differential equation of the inflaton
fluctuations. We may conclude that they acquire the amplitude
$\delta\sigma_*\simeq H_*/2\pi$. On the other hand, outside of the
horizon, the fluctuations obey the same differential equation as
the unperturbed curvaton field and then, we expect that they
remain constant during inflation. The Bardeen parameter,
$P_\zeta$, whose observed value is about $2\times 10^{-9}$
\cite{WMAP3}, allows us to determine the value of the curvaton
field, $\sigma_*$, in terms of the parameters $A$ and $\beta$. At
the time when the decay of the curvaton field occurs the Bardeen
parameter becomes \cite{ref1u}
\begin{equation}
P_\zeta\simeq \frac{1}{9\pi^2}\frac{H_*^2}{\sigma_*^2}.
\label{pafter}
\end{equation}
The spectrum of fluctuations is automatically gaussian for
$\sigma_*^2\gg H_*^2/4\pi^2$, and is independent of
$\Gamma_\sigma$ \cite{ref1u}. This feature will simplify the
analysis in the space parameter of our model.

From expression (\ref{pafter}), we may write
\begin{equation}
A=\left[\frac{27\pi^2}{48}\;\sigma_*^2\;(\beta+4)^2\;P_{\zeta}\left(\frac{\beta+1}{2}-N_*\right)^{\beta/2}
\right]^{\frac{2}{\beta+4}}, \label{18}
\end{equation}
where
\begin{equation}
N_*=\int_{t_*}^{t_e}\;H(t')\,dt'=\frac{1}{2\beta}\;(\phi_e^2-\phi_*^2),
\end{equation}
defines  the number of the e-folds corresponding to the
cosmological scales, i.e. the number of remaining inflationary
e-folds at the time when the cosmological scale exits the horizon.
Note that the parameter $\beta$ satisfies $\beta>2\,N_*-1$ or
equivalently $4\,(2N_*+3)^{-1}>f$.

Now, the constraint given by Eq. (\ref{one}) becomes
\begin{equation}
m^2 <
9\;\pi^2\;\sigma_*^2\;P_\zeta\;\left(1-\frac{2\;N_*}{\beta+1}\right)^{\beta/2},\label{21}
\end{equation}
and with the help of Eqs.(\ref{gamm1}) and (\ref{21}) we may write
\begin{equation}
\Gamma_\sigma <
\frac{\pi}{2}\;\sigma_*^3\;P_\zeta^{1/2}\;\left(1-\frac{2\;N_*}{\beta+1}\right)^{\beta/4},\label{ww}
\end{equation}
which gives an upper limits on $\Gamma_\sigma$ when the curvaton
field decays after domination.

On the other hand,  we give  the constraints on the parameters $A$
and $\beta$ by using the Big Bang Nucleosynthesis (BBN)
temperature $T_{BBN}$. We know that reheating occurs before the
BBN, where the temperature is of the order of $T_{BBN}\sim
10^{-22}$ (in unit of $m_p$),  and thus the reheating temperature
has to satisfies the inequality    $T_{reh}>T_{BBN}$. By  using
that $T_{reh}\sim \Gamma_\sigma^{1/2}\,>\,T_{BBN}$ we obtain the
constraint
\begin{equation}
H_{*}^2=16 \left( \frac{A}{4+\beta}\right)^2 \left
[\frac{2\,A}{\beta+1-2\,N_*} \right]^{\beta/2}
>\left(540\,\pi^2\right)^{2/3}\,P_\zeta^{2/3}\,T_{BBN}^{4/3}\sim
10^{-33},\label{c}
\end{equation}
where we have taken the scalar spectral index $n_s$ closed to one,
and therefore $m\leq 0.1\,H_*$. Note that Eq.(\ref{c}) is similar
to that described in ref.\cite{BuDi}. Note also this constrain
gives a lower limit for the parameters $A$ and $\beta$. On the
other hand, following the same ref.\cite{BuDi}, we could write an
upper limit for the Hubble parameter $H_*$, which satisfies the
inequality $H_*\leq 10^{-5}$.

\begin{figure}[th]
\includegraphics[width=9.0in,angle=0,clip=true]{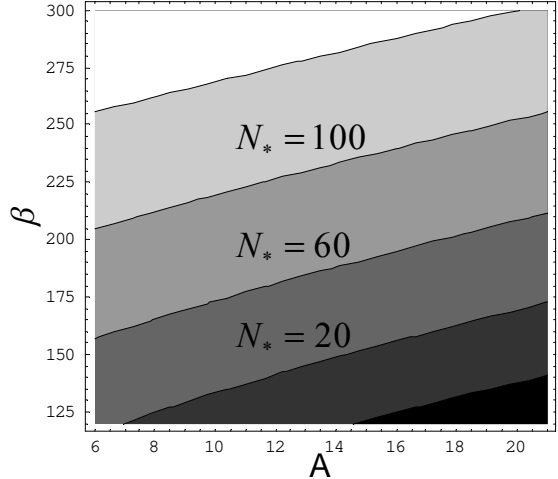}
\vspace{-10cm}\caption{ Contour plot for the number $N_*$ of
e-folds as a function of the parameters $\beta$ and $A$, fitted
from the lower limit of the BBN temperature (see Eq.(\ref{c})).
Lower values: the $N_*$ parameters correspond to darker regions
and the contour levels are separated by the quantity $\Delta
N_*=40$. \label{cou}}
\end{figure}

\begin{figure}[th]
\includegraphics[width=9.0in,angle=0,clip=true]{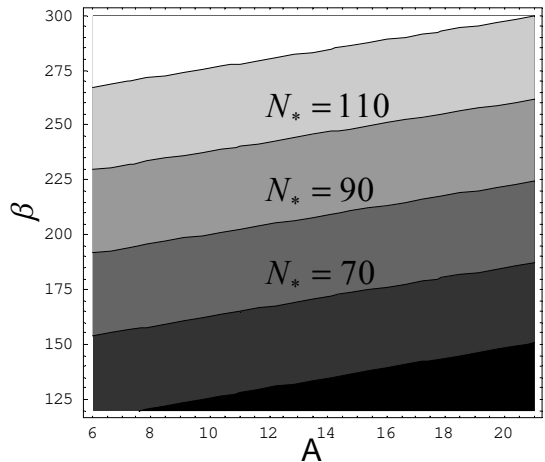}
\vspace{-10cm}\caption{ Contour plot for the number $N_*$ of
e-folds as a function of the parameters $\beta$ and $A$, fitted
from the upper limit, $H_*\leq 10^{-5}$. Lower values the $N_*$
parameters correspond to darker regions and the contour levels are
separated by the quantity $\Delta N_*=20$. \label{cou2}}
\end{figure}

In  Fig.\ref{cou} we plot contours curves corresponding to the
same number of e-folds, $N_*$, and different combinations of the
$\beta$ and $A$ parameters by fitting Eq.(\ref{c}) in its lower
limit. Here, we have taken $T_{BBN}\sim 10^{-22}$ (in units of
$m_p$). From this plot, given a value of $N_*$, one can therefore
constrain the values of $\beta$ and $A$ parameters. For instant,
over the line $N_*=60$ we could extract the values $\beta= 196$
and $A=18$ and so for others values of these parameters. A similar
graph is obtained when the upper limit, $H_* \leq 10^{-5}$, is
used, except that the contour lines get bigger values for the
$N_*$ parameter (see Fig.\ref{cou2}).

\section{Curvaton Decay Before Domination\label{sectv}}

For the second scenario, we assume that the decay of the curvaton
field happens before this dominates the cosmological expansion. In
this way, we need that the curvaton decays before its energy
density becomes greater than the inflaton one. Additionally, the
mass of the curvaton is non-negligible when compared with the
Hubble expansion rate $H$, so that we could use
$\rho_\sigma\propto\;a^{-3}$. We may say that the curvaton field
decays at a time when $\Gamma_\sigma =H(a_d)=H_d$ and then from
Eq. (\ref{h}) we get

\begin{equation}
\Gamma_\sigma=\;H_d=\;H_k\;\left(\frac{a_k}{a_d}\right)^3,
\label{Gamm}
\end{equation}
where ` d' stands for quantities at the time when the curvaton
decays.

If we allow the decaying of the curvaton field  after its mass
becomes important, (so that $\Gamma_\sigma<m$) and before that the
curvaton field dominates the expansion of the universe (i.e.,
$\Gamma_\sigma>H_{eq}$), we may write a new constraint, given by

\begin{equation}
\frac{\sigma_*^2}{6}<\frac{\Gamma_\sigma}{m}<1, \label{gamm2}
\end{equation}
which results in being  the same as that obtained in
ref.\cite{ureña}.

In the second scenario, the curvaton decays at the time when
$\rho_\sigma<\rho_\phi$. If we define the $r_d$ parameter as the
ratio between the curvaton and the inflaton energy densities,
evaluated at the time in which the curvaton decay occurs, i.e. at
 $a=a_d$ and for $r_d\ll 1$ the Bardeen parameter results \cite{ref1u,L1L2}

\begin{equation}
P_\zeta\simeq \frac{r_d^2}{16\pi^2}\frac{H_*^2}{\sigma_*^2}.
\label{pbefore}
\end{equation}

Defining $r_d=\left.\frac{\rho_\sigma}{\rho_\phi}\right|_{a=a_d}
$, from which we get that $ r_d=\frac{m^2\;\sigma_*^2\; a_m^3
\;a_{d}^3}{6\;H_k^2\; a_k^6} $, where we have used
$\rho_\sigma(a)= \frac{m^2\sigma_*^2}{2}\left (
\frac{a_k}{a}\right )^3$ and $\rho_\phi(a)=\rho_\phi^k\left (
\frac{a_k}{a}\right )^6$, and using expressions (\ref{mh}) and
(\ref{Gamm}) we obtain
\begin{eqnarray}
r_d=\;\frac{m\;\sigma_*^2}{6\;\Gamma_\sigma}\label{rd}.
\end{eqnarray}

From expressions (\ref{pbefore}) and (\ref{rd}) we find that $
\sigma_*^2=576\;\pi^2\;\frac{P_\zeta}{m^2}
\frac{\Gamma_\sigma^2}{H_*^2}\,$ and using that
\begin{equation}
 H^2_*=\frac{V_*}{3}= 16 \left(
\frac{A}{4+\beta}\right)^2 \left [\frac{2\,A}{\beta+1-2\,N_*}
\right]^{\beta/2} \label{h2}, \end{equation} we get
\begin{equation}
\sigma_*^2=36\pi^2\frac{P_\zeta\;\Gamma_\sigma^2\;(\beta+4)^2}{m^2\;A^2}
\left[\frac{\beta+1-2\,N_*}{2\,A}\right]^{\beta/2}.
\end{equation}
Thus, expression (\ref{gamm2}) becomes $\ds
18\pi^2\frac{P_\zeta\;\Gamma_\sigma^2\;(\beta+4)^2}{m^2\;A^2}
\left[\frac{\beta+1-2\,N_*}{2A}\right]^{\beta/2}<\frac{\Gamma_\sigma}{m}<1$,
from which we could write the inequality
\begin{equation}
\Gamma_\sigma<\frac{1}{18\pi^2}\frac{m\;A^2}{P_\zeta(\beta+4)^2}
\left[\frac{2\,A}{\beta+1-2\,N_*}\right]^{\beta/2}\label{37}.
\end{equation}

We see that this inequality for $\Gamma_\sigma$ depends on the
free parameters, $A$ and $\beta$, characteristic of the
intermediate inflationary universe model.

Finally, we derive a constraint for the parameters $A$ and $\beta$
by using the  BBN temperature $T_{BBN}$. Since, the reheating
temperature satisfies the bound  $T_{reh}>T_{BBN}$, and also
$\Gamma_\sigma\,>\,T^2_{BBN}$ we get
\begin{equation}
H_{*}^2=16 \left( \frac{A}{4+\beta}\right)^2 \left
[\frac{2\,A}{\beta+1-2\,N_*} \right]^{\beta/2}
>\left(960\,\pi^2\right)^{2/3}\,P_\zeta^{2/3}\,T_{BBN}^{4/3}\sim
10^{-33},\label{c3}
\end{equation}
where, as before, we have used the scalar spectral index $n_s$
closed to one. Note that this constrain  is similar to that
obtained when the curvaton field decays before domination, as
expressed by Eq.(\ref{c}).

\section{Constraints from gravitational waves\label{sectgw}}

In the same way that we have a constraint for $\Gamma_\sigma$
parameter, we could restrict the value of the curvaton mass, but
now  using tensor perturbations. In this kind of model the
corresponding gravitational wave amplitude   can be written as
\cite{Staro2}
\begin{equation}
h_{GW}\simeq\,C_1\,H_*,\label{po}
\end{equation}
where $C_1$ is an arbitrary constant.

Note that in this case we could take $H\ll 10^{-5}$ \cite{Dimo},
meaning that inflation may take place at an energy scale smaller
than the grand unification. In this way, this is an advantage of
the curvaton approach when compared with the single inflaton field
scenario.

Now, from the approximated Friedmann Eqs. we have
 $H^2_*=V_*/3$, and thus we may write for the gravitational  wave
 amplitude
\begin{equation}
h_{GW}^2\simeq\,16\;C_1^2\;\left(\frac{A}{\beta+4}\right)^2
\;\left(\frac{2A}{\beta+1-2\,N_*}\right)^{\beta/2}
  ,\label{gw}
\end{equation}
where we have used Eq.(\ref{h2}).

From Eqs.(\ref{one}) and (\ref{gw}) we get the inequality
\begin{equation}
m^2
<\,\frac{h_{GW}^2}{C_1^2}\,\left(\frac{\beta+1-2\,N_*}{\beta+1}\right)^{\beta/2}
,\label{m2}
\end{equation}
which gives an upper limit for the curvaton mass.

If we consider that $h_{GW}$ of the order of $10^{-5}$, and we
take $C_1\simeq 10^{-5}$ and $\beta = 250 $, and if we take the
number of e-fold to be $N_*= 60$, then we find that the above
equation gives the following upper limit for the curvaton mass (in
units of $m_p$)
\begin{equation}
m\,<\;10^{-18}.\label{mm}
\end{equation}
This value is closed to that considered in ref.\cite{BuDi}.

Since after inflation the  inflaton field  follows an equation of
state which is almost stiff the spectrum of relic gravitons
presents a characteristic in which the slope grows with the
frequency (spike) for models that re-enter  the horizon during
this epoch. This means that at high frequencies the spectrum forms
a spike instead of being of flat as in the case of radiation
dominated universe\cite{Gi}. Therefore, high frequency gravitons
re-entering the horizon during the kinetic epoch may disrupt BBN
by increasing the Hubble parameter. This problem can be  avoided
if  the following constraint is required on the density fraction
of the gravitational wave\cite{Dimopoulus}
\begin{equation}
I\equiv\,h^2\,\int_{k_{BBN}}^{k_{*}}\;\Omega_{GW}(k)\,d\ln\,k\simeq\,2\,
h^2\,\epsilon\,\Omega_\gamma(k_0)\,
h_{GW}^2\left(\frac{H_*}{ \widetilde{H}}\right)^{2/3}\leq\,2\times
10^{-6},\label{I}
\end{equation}
where $\Omega_{GW}(k)$ is the density fraction of the
gravitational wave with physical momentum $k$, $k_{BBN}$ is the
physical momentum  corresponding to the horizon at BBN,
$\Omega_\gamma(k_0)=2.6\times 10^{-5}h^{-2}$ is the density
fraction of the radiation at present on horizon scales. Here,
$\epsilon\sim 10^{-2}$ and  $h=0.73$ is the Hubble constant in
which $H_0$ is in units of 100 km/sec/Mpc.  The parameter
$\widetilde{H}$ represents either $\widetilde{H}=H_{eq}$, when the
curvaton decays after domination, or $\widetilde{H}=H_{d}$, if the
curvaton decays before domination.

For the first  scenario, the decay of the curvaton field happens
after that  this field dominates the cosmological expansion. In
this way, the constraint on the density fraction of the
gravitational wave, expressed by Eq.(\ref{I}), becomes
\begin{equation}
\frac{m}{\sigma_*^2}\gtrsim\,\left(\frac{\,P_{\zeta}}{4\times10^{5}}\right)^2\sim
10^{-28},\label{II}
\end{equation}
where we have used  expressions (\ref{heq}), (\ref{po}) and
$C_1\sim10^{-5}$.

When the decay of the curvaton field happens before that this
field dominates, the constraint on the density fraction of the
gravitational wave given by Eq.(\ref{I}), becomes
\begin{equation}
\frac{m^2\;\sigma_*^2}{\Gamma_\sigma^{1/4}}\gtrsim\,6\times10^{-5}\,P_{\zeta}\sim
10^{-13},\label{I1}
\end{equation}
where we have used  Eqs.(\ref{Gamm}) and (\ref{pbefore}).

Another set of bounds could be put forward by considering   the
decay rate of the curvaton field $\sigma$, which, in a very
particular case could be consider to be  $\Gamma_\sigma=g^2\,m$
ref.\cite{BuDi}, where $g$ is the coupling of the curvaton to its
decay products. The allowed range for the coupling constant in
this case becomes given by the expression
\begin{equation}
\mbox{max}\left(\frac{T_{BBN}}{m^{1/2}},m\right)\lesssim
g\,\lesssim\,\mbox{min}\left(1,
\frac{m\,\sigma^3}{T_{BBN}^{2}}\right), \label{g}
\end{equation}
where the inequality  $m\lesssim g$  is due to gravitation decay.
For the curvaton decays before domination and $T_{reh}>T_{BBN}$
this constrain gives an upper limit given by
$g<m\sigma_*^3/T_{BBN}^2$, and  when the curvaton decays after
domination, a lower limit is obtain given by
$T_{BBN}\,m^{-1/2}<g$.

\section{Conclusions \label{conclu}}

We have introduced the curvaton field in the intermediate
inflationary universe model. We have describe the curvaton
reheating  in which we  considered two cases. In the first case
the curvaton dominates the universe before it decays. Here, we
have arrived to Eq.(\ref{18}), which represents an interesting
constraint for the $A$ parameter  that appears in the scale factor
(see Eq.(\ref{at})). In the second case the curvaton decays before
domination. Here, we have found a restriction for the
$\Gamma_\sigma$ parameter, as shown by Eq.(\ref{37}).

In the context of the curvaton scenario, reheating does occur at
the time when the curvaton decays, but only in the period when the
curvaton dominates. In contrast, if the curvaton decays before its
density dominates the universe, reheating occurs when the
radiation due to the curvaton decay manages to dominate the
universe.

During the epoch in which the curvaton decays after its dominates
($\rho_\sigma >\rho_\phi$), the reheating temperature is the order
of $3\times 10^{-21}$ (in units of $m_p$), since the decay
parameter $\Gamma_{\sigma}\propto\,T_{rh}^2$, where $T_{rh}$
represents the reheating  temperature. Here, we have used Eq.
(\ref{ww}), with $\sigma_*\sim 10^{-9}$, $N_*=60$
 and $\beta=250$.  We should note that this  upper limit for  $T_{rh}$,
could  be modified.  Now, by using Eqs. (\ref{mm}) and (\ref{g})
we obtain that $10^{-13}\lesssim g\lesssim 10^{-1}$, and from
Eq.(\ref{gamm2}) we get that $H_d/\Gamma_\sigma\sim
10^{-18}g^{-2}$. If the decay of the curvaton field happens after
domination, then it is found the range  $10^{-13}\lesssim g
\lesssim 10^{-9}$. In this way since $T_{reh}\sim g\,m^{-1/2}$,
the allowed range for the reheating temperature becomes $10^{-22}
\lesssim T_{reh}\lesssim 10^{-18}$ (in units of $m_p$). Note that,
the bounds given by Eqs.(\ref{II}) and (\ref{I1}) may truncate
further the range for the Hubble parameter expressed by
$10^{-17}\leq H_*\leq 10^{-5}$ and Eq.(\ref{g}).

 Now,
let us choose $\sigma_*\sim 1$ (following ref.\cite{BuDi}),
$N_*=60$  and $\beta=250$.  From expression  (\ref{g}) the ranger
for
 $g$ becomes $10^{-13}\lesssim g \lesssim 1$, and since
  $H_d/\Gamma_\sigma\sim \sigma_*^2\,g^{-2}\gtrsim 1$, the curvaton
  decay it produced at or after domination. Therefore, with
$T_{reh}\sim g\,m^{-1/2}$,  the allowed ranger for the reheating
temperature becomes given by $10^{-22} \lesssim T_{reh} \lesssim
10^{-9}$ (in units of $m_p$). The constraints on the density
fraction of Gravitational Waves suggest $g\sim 1$\cite{BuDi}. In
this case,we obtain that the reheating temperature becomes of the
order of $ T_{rh}\sim 10^{-9}$ (in units of $m_p$), which
seriously challenges gravitino constraints \cite{rtref}.

\begin{acknowledgments}
S. d. C. was supported by Comision Nacional de Ciencias y
Tecnolog\'{\i}a through FONDECYT grants N$^0$ 1040624, N$^0$
1051086 and N$^0$ 1070306, and also from UCV-DGIP N$^0$
123.787/2007. R. H. was supported by the ``Programa Bicentenario
de Ciencia y Tecnolog\'{\i}a" through the Grant ``Inserci\'on de
Investigadores Postdoctorales en la Academia" \mbox {N$^0$
PSD/06}.

\end{acknowledgments}


\end{document}